\documentclass{aa}  
\usepackage{natbib}
\usepackage{graphicx}
\usepackage{txfonts}
\usepackage{xcolor}

\bibpunct{(}{)}{;}{a}{}{,} 
\definecolor{links}{rgb}{0, 0, 255}
\usepackage[colorlinks=true,allcolors=links]{hyperref}
\newcommand{\mpchi}{\,h^{-1}{\rm {Mpc}}}

\newcommand{\msun}{M_{\sun}}

\newcommand{\hi}{H{~\sc i}\xspace}
\newcommand{\hj}{\textrm{H \textsc{i}}}

\begin{document} 

\title{\textsc{NeutralUniverseMachine}: How filaments and dark-matter haloes influence the galaxy's cold gas content}

\author{
Wenlin Ma\inst{\ref{inst1},\ref{inst2}}
\and Hong Guo\inst{\ref{inst1}}\fnmsep\thanks{Corresponding author: guohong@shao.ac.cn.}
\and Michael G. Jones\inst{\ref{inst3}} 
}

\institute{
Shanghai Astronomical Observatory, Chinese Academy of Sciences, Shanghai 200030, China.\label{inst1} 
\and University of Chinese Academy of Sciences, Beijing 100049, China.\label{inst2} 
\and Steward Observatory, University of Arizona, 933 N Cherry Ave., Tucson, AZ 85721, USA.\label{inst3}
}


 
  \abstract
   {}
   {We aim to investigate the influence of the distance to filaments and dark-matter haloes on galaxy cold-gas content in the empirical model \textsc{NeutralUniverseMachine} (NUM) and the hydrodynamical simulation IllustrisTNG.}
   {We used DisPerSE to identify cosmic web structures and calculate the distance of galaxies to filaments for both observations and models. We show the results of the \hi and $\rm H_2$ mass functions, \hi- and $\rm H_2$-halo-mass relations, \hi- and $\rm H_2$-stellar-mass relations for galaxies in the NUM model and IllustrisTNG with different distances to filaments and compare them with observational measurements. We also show the evolution of \hi\ and $\rm H_2$ mass densities at different distances to filament bins.}
   {We find that how filaments affect the \hi gas is generally less significant compared to the halo environment. There is a weak trend in the observations at $z=0$ that low-mass haloes lying closer to the filaments tend to have reduced \hi masses. However, this trend reverses for massive haloes with $\log(M_{\rm vir}/\msun)>12.5$. This behaviour is accurately reproduced in the NUM model due to the dependence of \hi gas on the halo formation time, but it does not appear in IllustrisTNG. The influence of filaments on the \hi gas becomes slightly weaker at higher redshifts and is only significant for galaxies that reside in massive haloes in the NUM model. Filaments have almost no impact on the H$_2$-stellar-mass relation in both models, confirming that H$_2$ is primarily determined by the galaxy stellar mass and star formation rate. }
   {}


   \maketitle
%

\section{Introduction}\label{sec:intro}

Galaxies in the Universe form the large-scale structure called the cosmic web \citep{Bond1996}. From high- to low-density regions, it consists of nodes (clusters), filaments, and voids. Filaments are intermediate density environments with long and thin structures. It can be clearly observed in the galaxy distributions from the redshift surveys; for example, the Sloan Digital Sky Survey \citep[SDSS;][]{Abazajian2009}, the Galaxy And Mass Assembly \citep[GAMA;][]{Driver2011}, and the 2MASS Redshift Survey \citep{Huchra2012}. $N$-body and hydrodynamical simulations can also reproduce these features \citep[e.g.][]{Springel2006, Vogelsberger2014, Nelson2019}.

Filaments account for roughly 40\% of the Universe's mass \citep{Aragon-Calvo2010} and encompass the majority of galaxies. Galaxies grow and develop within the cosmic web, and their characteristics are influenced by the large-scale environment, including their stellar mass, star formation rate (SFR), and morphology \citep[e.g.][]{Tanaka2004, Grutzbauch2011, Wang2018, Donnan2022}. Filaments also act as the main pathways of the cosmic web by transporting cold gas, particularly atomic and molecular hydrogen (\hi and $\rm H_2$), to nodes where massive clusters form. Several studies have shown that cold gas within galaxies interacts with the cosmic web through gas accretion, stripping, and other physical processes \citep{Mori2000, Quilis2000, Kawata2008, Danovich2012, Aragon2019}. 

The impact of the cosmic web on the cold-gas content of galaxies remains a topic of discussion. Some research suggests that the cosmic web can block the supply of cold gas to galaxies. For example, \cite{Odekon2018} employed the Arecibo Fast Legacy ALFA Survey (ALFALFA) \hi survey \citep{Giovanelli2005} and discovered that late-type galaxies lose \hi as they move into denser regions. Using data from REsolved Spectroscopy of a Local VolumE \citep[RESOLVE;][]{Eckert2015, Stark2016} and the Environmental COntext survey \citep[ECO;][]{Moffett2015}, \cite{Hoosain2024} found that galaxies located near filaments have reduced gas content, and more isolated galaxies become gas-deficient when closer to filaments.

Some studies argue that the cosmic web can replenish a galaxy's cold-gas content. Using the six-degree Field Galaxy Survey \citep{Jones2009} to select galaxies within filaments and the \hi Parkes all-sky survey \citep{Barnes2001} to measure \hi mass by spectral stacking, \cite{Kleiner2017} found that massive galaxies can acquire \hi through cold-gas accretion from filaments. Using colour-selected star-forming galaxies in the COSMOS field \citep{Scoville2007} at $z\sim0.37$ and stacking the \hi spectral from the MeerKAT International GigaHertz Tiered Extragalactic Exploration Survey \citep[MIGHTEE;][]{Jarvis2016}, \cite{Sinigaglia2024} found that galaxies in the high-density region have slightly higher \hi mass than those in the low-density region. Such results can be understood within the detachment model of \cite{Aragon2019}, which is related to the non-linear interaction between galaxies and filaments. Galaxies can accrete cold gas from the cosmic web and form stars when they move close to filaments until they first interact with the proto-cluster at the intersection of filaments. 

Alternatively, some studies support both situations when accounting for different galaxies. \cite{Kotecha2022} found that satellite galaxies within clusters and close to filaments tend to contain more cold gas than those further away from filaments, while outside clusters the cold-gas content of galaxies decreases when close to filaments in the Three Hundred project \citep{Cui2018}. \cite{Bulichi2024} showed that in SIMBA simulations \citep{Dave2019} the \hi fraction decreases in both central and satellite galaxies when they approach the filaments at $0<z<2$. 

In addition to the large-scale environment, the local environment also has non-negligible effects on galaxy evolution \citep{White1978, Blumenthal1984}. The impact of the large-scale structure on the cold-gas content of galaxies has been discussed in the literature, but local environmental factors may also play an important role. The cosmic web consists of dark-matter haloes and promotes the formation of haloes. The halo spin, shape, and formation time are correlated with the cosmic web \citep[e.g.][]{Sheth2004,Wang2011}. Meanwhile, galaxies reside in haloes. The stellar-mass assembly of galaxies depends primarily on the assembly of their host halo mass \citep{Behroozi2010, Tinker2013, Gu2016, Legrand2019}. \cite{Wangkuan2024} used the IllustrisTNG simulation and found that the distance to filaments or nodes of host haloes can account for a substantial part of the total secondary galaxy bias, emphasising the necessity of excluding halo mass effects when considering a large-scale environment. Galaxy properties depend on their host halo as well \citep[e.g.][]{Bethermin2014, Berti2017, Behroozi2019}. So, it is important to investigate whether the local environment or large-scale structure has a greater impact on the cold gas content of galaxies.

Hydrodynamical simulations, semi-analytic models, N-body simulations, and empirical models are effective ways to investigate the cold-gas content of galaxies within current galaxy formation models. Recently, \cite{Guo2023} proposed a new empirical model, \textsc{NeutralUniverseMachine}, to accurately reproduce the galaxy \hi and $\rm H_2$ content as well as their evolution. In this paper, we compare the \textsc{NeutralUniverseMachine} model and IllustrisTNG hydrodynamical simulations \citep{Nelson2018, Pillepich2018, Springel2018} with observational measurements to investigate how filaments affect cold-gas content and the impact that filaments and dark-matter haloes have on the cold-gas content of galaxies within current models.

This paper is organised as follows. We describe the data and how we calculated the distance to filaments in Section~\ref{sec:data}. The results  are given in Section~\ref{sec:result}, the discussion in Section~\ref{sec:discussion}, and the conclusion in Section~\ref{sec:conclusion}. Throughout the paper, the distance to the filament is in the comoving unit of $\mpchi$.

\section{Data and methods}\label{sec:data} 

\subsection{\textsc{NeutralUniverseMachine} model}\label{subsec: data NUM} 
The \textsc{NeutralUniverseMachine}\footnote{\url{https://halos.as.arizona.edu/UniverseMachine/DR1/Gas_Masses_NeutralUniverseMachine/}} (hereafter NUM) model \citep{Guo2023} is an empirical model that can self-consistently predict the evolution of neutral hydrogen (both \hi and $\rm H_2$) at $0<z<6$. It is based on \textsc{UniverseMachine} \citep{Behroozi2019}, which successfully reproduces observational galaxy properties, especially galaxy and halo assembly correlation, the quenched fraction, and star formation over $0<z<10$. As in UniverseMachine, the NUM model uses the Bolshoi–Planck N-body simulation \citep{Klypin2016} with a side length of $250~\mpchi$. It adopts a flat Lambda cold dark matter ($\Lambda$CDM) cosmology with $\Omega_{\rm m}= 0.307$, $h = 0.678$, $\Omega_{\rm b} = 0.048,$ and $\sigma_{\rm 8} = 0.823$. The mass resolution of dark-matter particles is $2.3 \times 10^{8}\msun$. The ROCKSTAR halo finder \citep{Behroozi2013b} is applied to define dark-matter haloes, and the CONSISTENT TREES algorithm \citep{Behroozi2013a} is used to construct halo merger trees.

In NUM, the \hi mass ($M_{\rm \hj}$) within a halo or sub-halo is determined as a function of its halo virial mass ($M_{\rm vir}$), halo formation time ($z_{\rm form}$), SFR, and redshift (z) as follows:
\begin{eqnarray}
    M_{\rm \hj} &=& \frac{\kappa M_{\rm vir}}{\mu ^{-\alpha} + \mu ^ \beta } (\frac{1+z}{1+z_{\rm form}})^\gamma (\frac{\rm SFR}{\rm SFR_{MS,obs}})^\lambda,       \label{eq:MHI}  \\
    \mu &=& M_{\rm vir}/M_{\rm crit}, \\
    \log \kappa &=& \kappa_{0} + \kappa_{1}z +\kappa_{2}z^2, \\
    \log M_{\rm crit} &=& M_0 + M_1z + M_2z^2,
\end{eqnarray}
where $\rm SFR_{MS,obs}$ is the star-forming main sequence defined in \cite{Behroozi2019}; and  $\kappa$, $\kappa_{1}$, $\kappa_{2}$, $M_0$, $M_1$, $M_2$, $\alpha$, $\beta$, $\gamma,$ and $\lambda$ are the model parameters.

The H$_2$ masses are determined following the definition of \cite{Tacconi2020} as 
\begin{eqnarray}
    M_{\rm H_2}&=&\zeta M_\ast ^ \nu (\frac{\rm SFR}{\rm SFR_{\rm MS,obs}})^\eta,  \label{eq:MH2} \\
    \log \zeta &=& \zeta_0 + \zeta_{1}\ln (1+z) + \zeta_2[\ln (1+z)]^2,
\end{eqnarray}
where $\zeta_0$, $\zeta_1$, $\zeta_2$, $\nu,$ and $\eta$ are also model parameters.

\subsection{IllustrisTNG simulation}\label{subsec: data TNG}
The IllustrisTNG project \citep{Nelson2018, Nelson2019} consists of a suite of cosmological hydrodynamical simulations run with the {\scriptsize AREPO} code \citep{Springel2010,Pakmor2016}. A flat $\Lambda$CDM cosmology is adopted, with parameters of $\Omega_m = 0.3089$, $\Omega_\Lambda = 0.6911$, $\Omega_b = 0.0486$, $\sigma_8 = 0.8159,$ and $h = 0.6774$. It contains three sets of different volumes -- TNG50, TNG100, TNG300 -- with box side lengths of roughly 50~Mpc, 100~Mpc, and 300~Mpc, respectively. In this paper, we focus on the TNG100 simulation, which has $2 \times 1820^3$ particles with a baryonic mass resolution of $1.4 \times 10^6~\msun$ and dark-matter resolution of $7.5 \times 10^6~\msun$.

In TNG100, galaxies and haloes are identified using the SUBFIND algorithm \citep{Springel2001}. The stellar mass of each galaxy is defined as the total mass of stellar particles within twice the stellar half-mass radius \citep{Nelson2019}. Star formation is regulated by a two-phase interstellar-medium model of \cite{Springel2003}, in which star formation develops when the density of the gas cell is greater than $n_{\rm H} = 0.106\,{\rm cm}^{-3}$. For \hi and $\rm H_2$ masses, we employed the post-processing framework described in \cite{Diemer2018}. They projected all gas cells in a galaxy onto a face-on grid and then used these projected quantities to estimate the molecular fraction $f_{\rm H_2}$ as a function of the average surface density, the UV field, and the metallicity. The \hi and $\rm H_2$ masses can then be obtained directly from the total neutral hydrogen mass using $f_{\rm H_2}$. To ensure a well-defined sample, the post-processing of \cite{Diemer2018} was only done for galaxies with a minimum stellar mass of $M_\ast>2\times10^8\msun$ or a minimum gas mass of $M_{\rm gas}>2\times10^8\msun$. They included five different methods to model the \hi/$\rm H_2$ transition, denoted as the L08 \citep{Leroy2008}, GK11 \citep{Gnedin2011}, K13 \citep{Krumholz2013}, GD14 \citep{Gnedin2014}, and S14 \citep{Sternberg2014} models. However, different models produce very similar results of the overall \hi and H$_2$ masses as shown in the appendix of \cite{Ma2022}. In this paper, we simply adopt the same choice of the K13 model as in \cite{Ma2022}.  In \cite{Krumholz2013}, the molecular fraction ($f_{\rm H_2}$) and the star formation rate would be affected by the interstellar radiation field and the cold gas density within the interstellar medium (ISM). $f_{\rm H_2}$ is defined as follows:
\begin{equation}
        f_{\rm H_2} =
        \begin{cases}
            1-3s/(4+s) & \text{if $s<2$} \\
            0          & \text{if $s \geq 2$}
        \end{cases},
    \label{eq:tng_fh2}
\end{equation}
where 
\begin{equation}
        s\equiv \rm{ \frac{\ln{(1+0.6\chi+0.01\chi^2)}}{0.6\tau_c}}   \label{eq:tng_s}
;\end{equation}
and
\begin{equation}
                \chi \equiv \rm{7.2U_{MW}( \frac{n_{CNM}}{10cm^{-3}})^{-1}} , \label{eq:tng_chi}
\end{equation}
where $\rm{U_{MW}}$ is UV background, $\rm{n_{CNM}}$ is cold-phase column density of the ISM, and $\tau_c$ is the optical depth of a cloud. 
We refer the reader to \cite{Diemer2018} for more details.

\subsection{Observational data}\label{subsec: obs} 
In this work, we adopted observational measurements of \hi and $\rm H_2$ -- including the \hi and $\rm H_2$ mass functions, the \hi -halo mass relation, and the \hi -stellar mass relation -- to compare with the results of NUM and TNG. The brief descriptions of these observations are as follows.

For mass functions, we used the \hi mass functions (HIMF) of \cite{Guo2023} and the $\rm H_2$ mass functions ($\rm H_2$MF) of \cite{Fletcher2021}. \cite{Guo2023} derived the HIMF from the final sample of ALFALFA \citep{Haynes2018} covering $\rm \sim 6900~deg^2$ in the redshift range of $0<z<0.05$. They used \hi targets above the 90\% completeness threshold to address the effect of incompleteness \citep{Jones2018}. \cite{Fletcher2021} measured the $\rm H_2$MF using 532 galaxies from the xCOLD GASS survey at $0.01<z<0.05$ \citep{Saintonge2017}.

The \hi-halo mass relation at $z\sim0$ was previously measured by \cite{Guo2020}. They selected haloes in the overlapping regions between ALFALFA and the galaxy group catalogue of SDSS DR7 \citep{Lim2017}. The average \hi masses for haloes in different mass bins are estimated by stacking the \hi spectra. In this work, we extended that of \cite{Guo2020} by further measuring the \hi-halo mass relation in different $d_{\rm fil}$ bins, using the same halo catalogue and stacking method.

The stacked \hi-stellar-mass relations for star-forming and quenched central galaxies at $z\sim0$ were measured by \cite{Guo2021} using the same set of SDSS galaxy samples as in \cite{Guo2020}. In this work, we measured the additional dependence of the \hi-stellar mass relation on $d_{\rm fil}$ for star-forming central galaxies, following the same method of \cite{Guo2021}. We also included the \hi measurements at $z\sim0.37$ from \cite{Sinigaglia2024}. They measured the stacked \hi mass of star-forming galaxies at $z\sim0.37$ using the MeerKAT International GigaHertz Tiered Extragalactic Exploration Large Survey Program \citep[MIGHTEE;][]{Jarvis2016} \hi survey. Using the overdensity of galaxy numbers ($\delta$), they classified high-density ($\delta>0$) and low-density ($\delta\leq0$) environments to investigate the \hi content of galaxies in different large-scale structure environments.

\subsection{Filament distance calculation}\label{subsec: filament distance} 

In this study, we used the Discrete Persistent Structures Extractor (DisPerSE) \citep{Sousbie2011a, Sousbie2011b} to detect filaments within the galaxy distribution. DisPerSE operates on the principles of Morse theory and facilitates the identification of nodes, filaments, walls, and voids. Nodes are defined as critical points where the gradient of the density field vanishes, whereas filaments correspond to the connections between these nodes. For this analysis, given the mass resolution, we focused on galaxies with $M_\ast > 10^{8.5}\msun$ to delineate the cosmic web. The details of the method are described in \cite{Wang2024}, which closely follows the procedures outlined in \cite{Galarraga2024}.

We measured the distance from a galaxy to the nearest filament point as $d_{\rm fil}$. A node represents the intersection of multiple filaments, which are characterised by a higher mass density and distinct dynamical environments that can influence the cold-gas distribution. We confirm that whether galaxies within $2\mpchi$ of nodes are included or excluded has a minimal impact on our conclusions. To improve the signal-to-noise ratio of both the observational data and the model outcomes, we chose not to exclude these galaxies close to the nodes, conversely to what was done by \cite{Wang2024}.

\section{Results}\label{sec:result} 
\subsection{Cold-gas distribution and mass function}\label{subsec:hih2mf} 
\begin{figure*}
\centering
    \includegraphics[width=\textwidth]{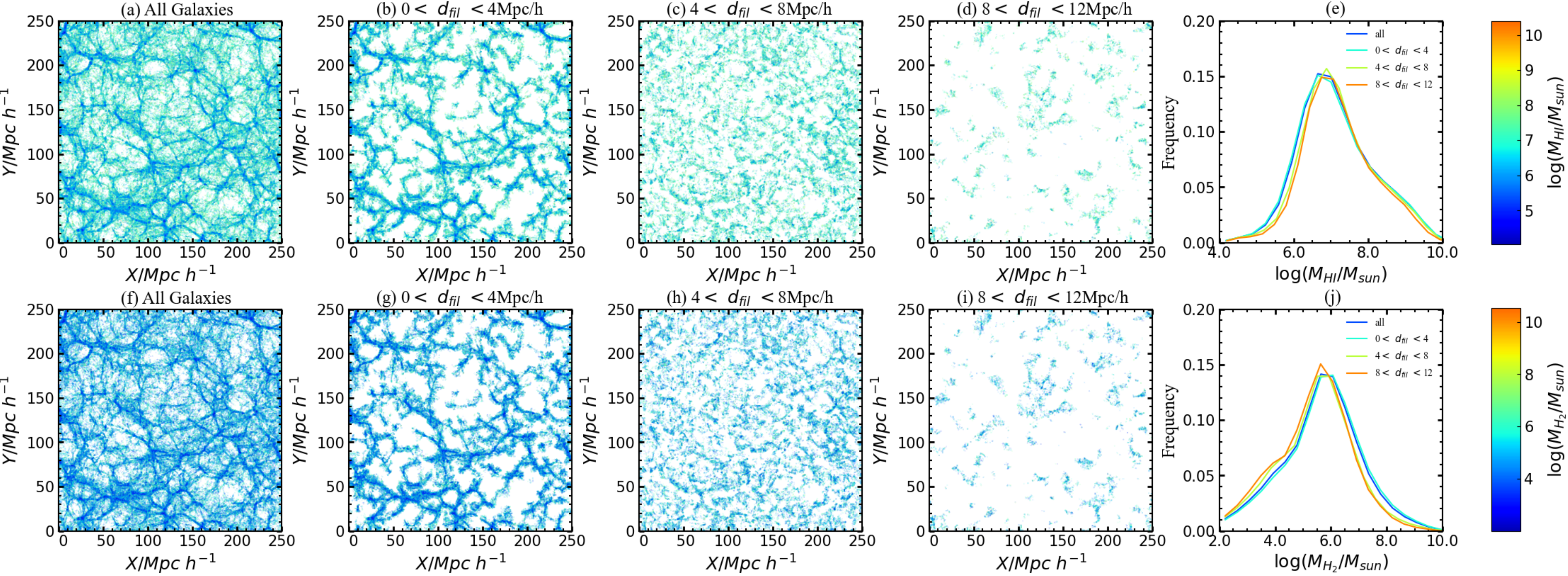}
    \caption{Galaxy distribution in NUM catalogue at $z = 0$. Top row: Panels (a) to (d), colour-coded by \hi mass, show the distribution for all galaxies (panel a), galaxies with with $0< d_{\rm fil}/\mpchi < 4$ (panel b), $4 < d_{\rm fil}/\mpchi< 8 $ (panel c), and $8< d_{\rm fil}/\mpchi< 12 $ (panel d), respectively. Panel (e) shows the \hi mass distribution for galaxies with different distances to the filament that are measured within the same thin slice as in panels (a)-(d); these are blue for all galaxies, cyan for $0 < d_{\rm fil}/\mpchi < 4$, green for $4 < d_{\rm fil}/\mpchi< 8 $, and orange for $8 < d_{\rm fil}/\mpchi < 12$. The bottom row is the same as top row, but colour-coded by $\rm H_2$ mass.}
    \label{fig_hih2map}
\end{figure*}

We display the galaxy distributions in the NUM model at $z = 0$ with a thin slice of $10\mpchi$ in Figure \ref{fig_hih2map}. The top and bottom panels are colour-coded by $M_{\rm \hj}$ and $M_{\rm H_2}$, respectively. The columns from left to right show the maps of all galaxies (panels a and f) and galaxies with $0 < d_{\rm fil}/\mpchi < 4$ (panels b and g), $4 < d_{\rm fil}/\mpchi < 8$ (panels c and h), and $8 < d_{\rm fil}/\mpchi < 12$ (panels d and i), respectively. Panels (e) and (j) illustrate the probability distributions of $M_{\rm \hj}$ and $M_{\rm H_2}$ for the four sub-samples. The distributions of $M_{\rm \hj}$ and $M_{\rm H_2}$ are largely consistent across the four sub-samples, suggesting that the distance to the filament may have a minimal impact on the probability distributions of the cold-gas mass. 

\begin{figure*}
    \centering
    \includegraphics[width=0.8\textwidth]{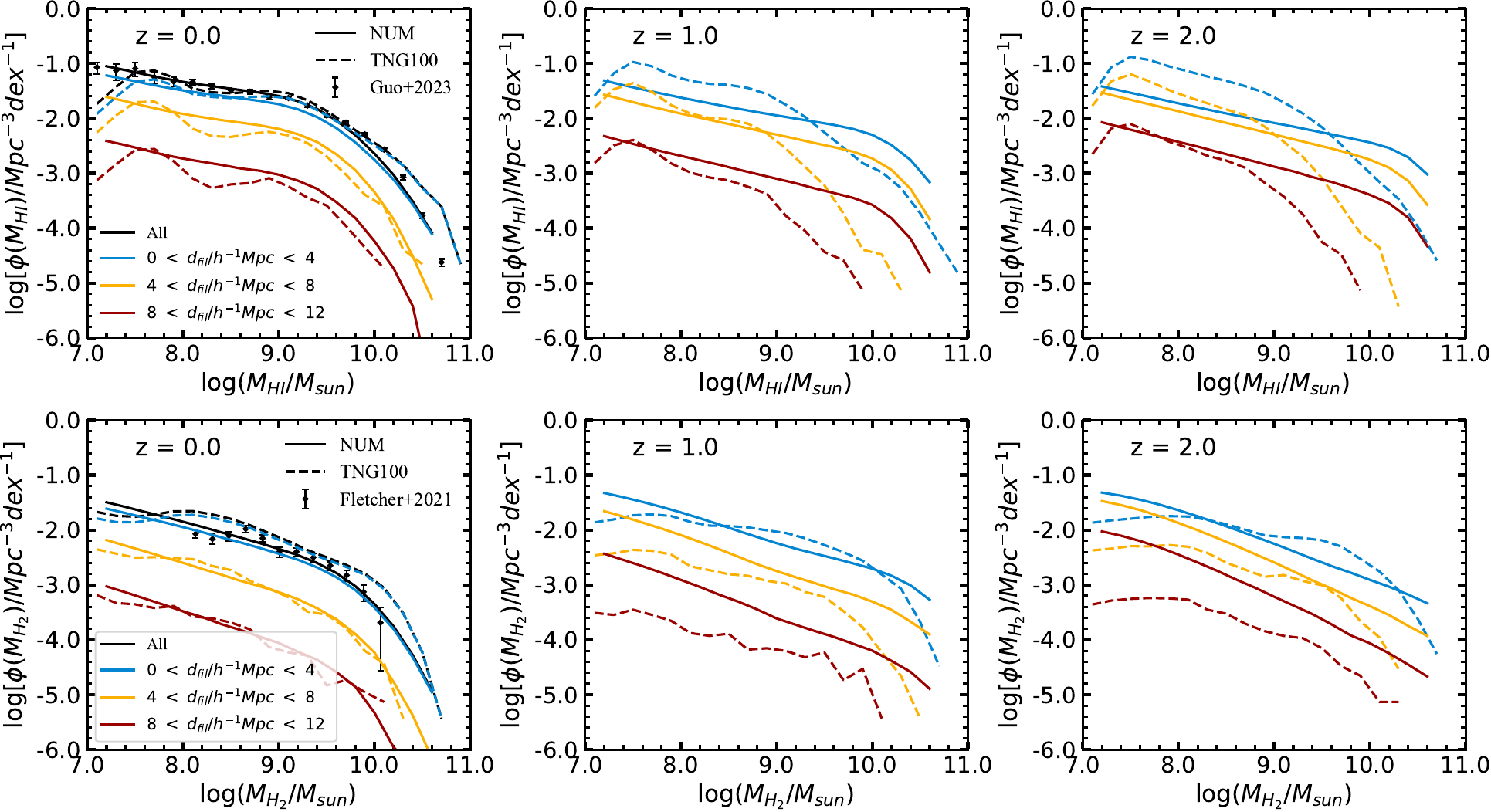}
    \caption{Comparisons of HIMF (top row) and $\rm H_2$MF (bottom row) between NUM model and TNG100 for $z = 0$ (left panels), $z = 1$ (middle panels), and $z = 2$ (right panels). We add observational measurements of HIMF in \cite{Guo2023} and $\rm H_2$MF in \cite{Fletcher2021} at $z = 0$ for reference in the top left panel and bottom left panel, respectively; these are shown as black dots with error bars. NUM model and TNG100 results are represented by solid and dashed lines, while the results of different distances to filaments are shown by different colours: black for all galaxies, blue for $\rm 0 < d_{fil}/\mpchi < 4$, orange for $\rm 4 < d_{fil}/\mpchi < 8$, and red for $\rm 8 < d_{fil}/\mpchi < 12$. }
    \label{fig_MF}
\end{figure*}

Figure~\ref{fig_MF} displays the HIMF (top panels) and the $\rm H_2$MF (bottom panels), respectively. The measurements for galaxies with varying $d_{\rm fil}$ are depicted as lines in different colours, as indicated in the legends. The predictions of the NUM model (solid lines) contrast with those of TNG100 (dotted lines) at $z = 0$ (left panels), $z=1$ (middle panels), and $z = 2$ (right panels). For reference, the observed HIMF of \cite{Guo2023} is included in the top left panel, and the $\rm H_2$MF of \cite{Fletcher2021} is included in the bottom left panel.

At $z = 0$, the NUM model closely matches the observed HIMF for all galaxies. TNG100 aligns well with the observations for $7.5 < \log (M_{\rm \hj }/\msun ) < 10$, but is restricted by the mass resolution at the lower mass end for both \hi and $\rm H_2$. TNG100 also slightly overestimates the count of galaxies with $\log (M_{\hj}/\msun)>10$. At $z = 1$ and $z = 2$, the shapes of HIMF and H$_2$ MF in TNG100 differ significantly from those predicted by the NUM model. Compared to the NUM model, TNG100 generally forecasts more galaxies with low $M_{\hj}$ and considerably fewer galaxies with high $M_{\hj}$. The H$_2$MF predictions of NUM and TNG100 are roughly consistent with each other at massive ends. However, it should be noted that for both \hi and $\rm H_2$ mass functions at all redshifts, galaxies with varying $d_{\rm fil}$ have very similar shapes in both NUM and TNG100. Our result is consistent with those of \cite{Moorman2014} and \cite{Jones2016}, which also found only slight changes of slope and knee mass in the HIMF in different large-scale environments. The proximity to filaments mainly affects the overall amplitude of the cold gas mass functions, rather than their shapes. In other words, there are more galaxies closer to the filaments, but the relative distribution of cold gas within galaxies at a fixed $d_{\rm fil}$ bin remains unaffected. This is consistent with the findings shown in the right panels of Figure~\ref{fig_hih2map}.

\subsection{The role of halo environment}\label{subsec:hih2mh} 
\begin{figure*}
    \centering
    \includegraphics[width=0.8\textwidth]{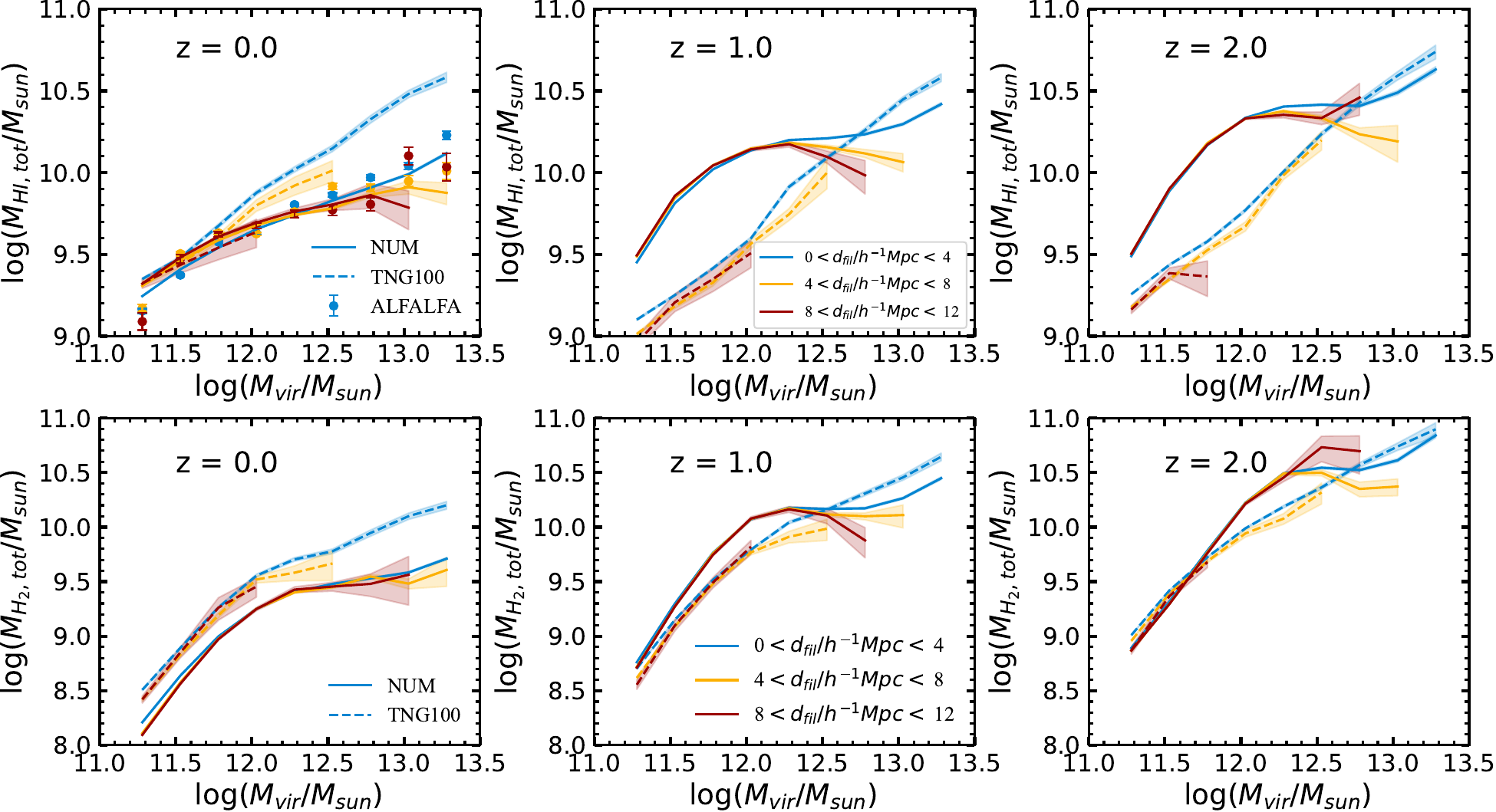}
    \caption{Comparisons of \hi\ - (top row) and $\rm H_2$- (bottom row) halo-mass relations between TNG100 and NUM for $z = 0$ (left panels), $z = 1$ (middle panels), and $z = 2$ (right panels). NUM and TNG100 are represented by solid and dashed lines, respectively, with shaded areas showing the bootstrapping errors; while the results of different distances to filaments are shown as different colours: blue for $\rm 0 < d_{\rm fil}/\mpchi < 4$, orange for $\rm 4 < d_{\rm fil}/\mpchi < 8$, and red for $\rm 8 < d_{\rm fil}/\mpchi < 12$. We add observational measurements in \cite{Guo2020} at $z=0$ as filled circles with error bars in the top left panel. }
    \label{fig_MH}
\end{figure*}
It has been well established that the \hi mass of a galaxy is strongly dependent on the properties of its host halo, including the halo mass \citep{Guo2020, Dev2023, Rhee2023} and the halo-formation time \citep{Guo2017, Stiskalek2021}, as depicted in the NUM model \citep{Guo2023}. Therefore, it is crucial to investigate the individual effects of the local halo environment and the cosmic web on the cold-gas content of galaxies. To achieve this, we analysed the \hi - and $\rm H_2$ -halo mass relations for galaxies with different $d_{\rm fil}$ in the top and bottom panels of Figure~\ref{fig_MH}, respectively. We compare NUM (solid lines) with TNG100 (dashed lines) in $z = 0$ (left panels), $z=1$ (middle panels), and $z = 2$ (right panels). The errors for each model are estimated with the bootstrapping method, shown as the shaded areas.

Extending the work of \cite{Guo2020}, we show the ALFALFA measurements at $z = 0$ as filled circles with error bars in the top left panel of Figure~\ref{fig_MH}. For both observations and models,  $M_{\rm HI,tot}$ and $M_{\rm H_2,tot}$ are the average \hi and $\rm H_2$ masses in a given halo-mass bin, which are the values of dividing the total \hi and $\rm H_2$ masses in each $M_{\rm vir}$ bin by the number of parent haloes.

In the ALFALFA data, there is a subtle trend whereby the \hi-halo-mass relation with varying $d_{\rm fil}$ transitions at $\log (M_{\rm vir}/\msun) \sim 12.5$. At $z = 0$, haloes closer to the filaments exhibit lower $M_{\rm HI,tot}$ for $\log (M_{\rm vir}/\msun)<12.5$ and higher $M_{\rm HI,tot}$ for more massive haloes. This pattern is accurately replicated in the NUM model. For $z \geq 1$, this switching dependence on $M_{\rm vir}$ in NUM becomes significantly weaker, but the influence of filaments on $M_{\hj,{\rm tot}}$ intensifies for massive haloes. The origin of the switch at $z=0$ is discussed in Section~\ref{subsec:dis_tranision}. In observational measurements and models, we do not exclude the haloes that reside close to the nodes (i.e. clusters) when measuring the cold-gas masses for $0 < d_{\rm fil}/\mpchi < 4$. This causes a strong increase of $M_{\hj,{\rm tot}}$ for $\log(M_{\rm vir}/\msun)>13$ in all redshifts, which is not seen in the other two $d_{\rm fil}$ bins. We verify that when the haloes with $d_{\rm node}<2\mpchi$ are removed, $M_{\hj,{\rm tot}}$ in NUM will become flat at the massive end. 

TNG100 displays markedly different trends. Haloes closer to the filaments consistently have higher $M_{\rm HI,tot}$, with values considerably higher than those observed for $\log (M_{\rm vir}/\msun)>12$ at $z=0$. As the redshift increases, these differences decrease, indicating a reduced impact of the filaments. At $z>1$, TNG100 predicts much smaller total \hi masses than NUM for low-mass haloes of $\log(M_{\rm vir}/\msun)<12.7$. This is related to the increase in the cosmic \hi density from $z=0$ to $z=1$ in NUM to match the observed trend \citep{Guo2023}, as well as the need to fit the \hi-stellar-mass relation at $z=1$ \citep{Chowdhury2022}. However, in TNG100, the cosmic \hi density remains roughly constant from $z=0$ to $z=4$ \citep{Villaescusa-Navarro2018}, resulting in weak evolution in the \hi-halo-mass relation. Across all redshifts, TNG100 produces few massive haloes distant from the filaments, as illustrated by the short red and orange lines.

Similarly, in the bottom panels of Figure~\ref{fig_MH}, we show the $\rm H_2$-halo-mass relations at different redshifts. The behaviour of $M_{\rm H_2, tot}$ is similar to that of $M_{\hj,{\rm tot}}$ in the upper panels. The dependence of $M_{\rm H_2, tot}$ on $d_{\rm fil}$ is slightly weaker. In NUM, the effect of $d_{\rm fil}$ is only seen in massive haloes of $\log(M_{\rm vir}/\msun)>12.5$ at high redshifts. As shown in Section~\ref{sec:data}, in the NUM model, $M_{\rm H_2}$, is directly determined by $M_\ast$ and SFR, and thus it is less affected by the large-scale environment. We note that TNG100 predicts a consistently higher $\rm H_2$-halo-mass relation than NUM at $z=0$. This is related to the overestimation of H2MF shown in the bottom left panel of Figure~\ref{fig_MF}. At higher redshifts, the discrepancies between the predictions of TNG100 and NUM become smaller. Both models show similar total H$_2$ masses for low-mass haloes. 

\subsection{\hi-stellar-mass relation of star-forming central galaxies}\label{subsec:hih2ms} 

\begin{figure*}
    \centering
    \includegraphics[width=\textwidth]{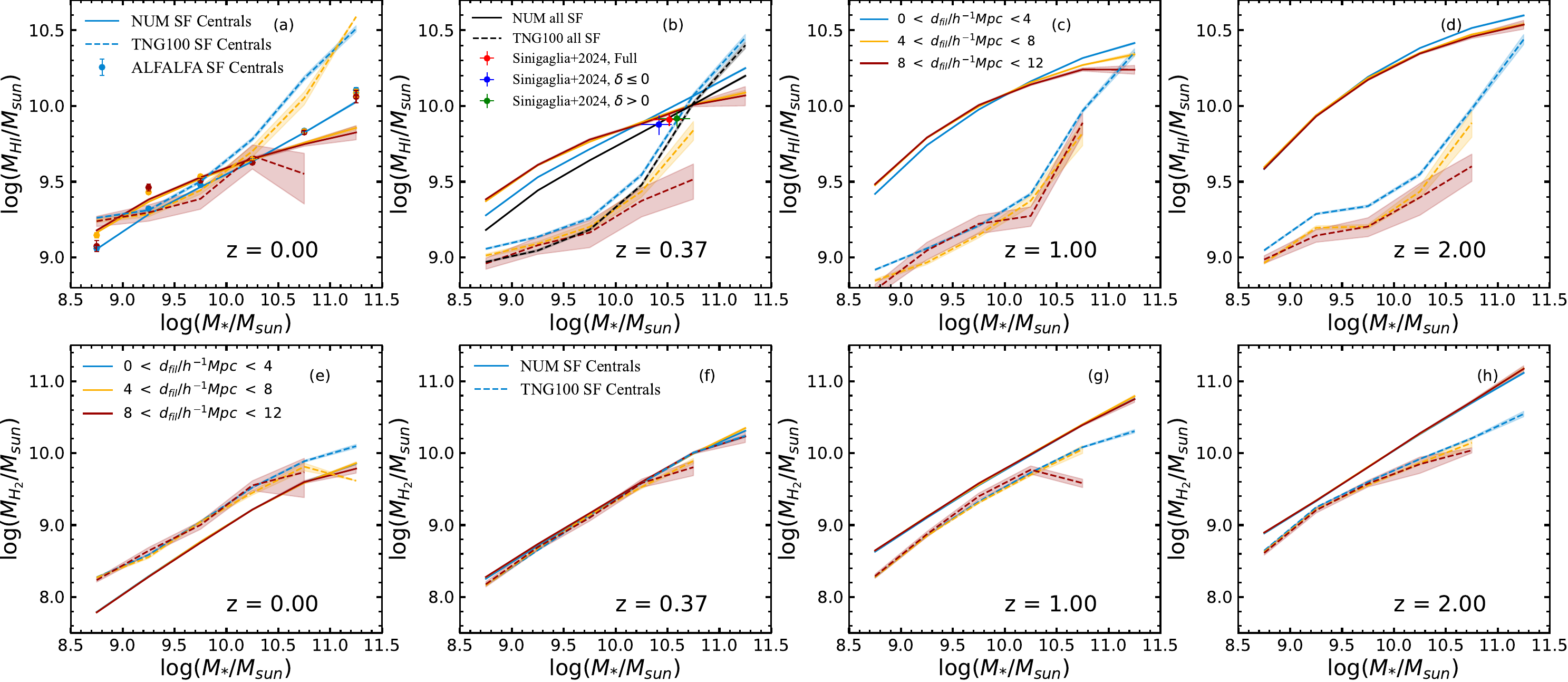}
    \caption{Comparisons of \hi\ - (top row) and $\rm H_2$- (bottom row) stellar-mass relations for star-forming (SF) central galaxies between NUM model and TNG100 for $z = 0$, $z = 0.37$, $z = 1$, and $z = 2$ from left to right panels. NUM and TNG100 results are shown as solid and dashed lines. The bootstrap errors are shown in the shaded area. As TNG100 does not have data for $z = 0.37$, we use $z = 0.5$ instead in panels (b) and (f). We show observational measurements in \cite{Guo2021} as filled circles with error bars for $z = 0$ in panel (a). In panel (b), we adopt the full galaxy sample, galaxies with $\delta \leq 0$, and those with $\delta > 0$ in \cite{Sinigaglia2024} as red, blue, and green circles with error bars, respectively. We add the relation of all star-forming galaxies in the NUM model and TNG100 in panel (b) as black solid and dashed lines for a fair comparison with \cite{Sinigaglia2024}. The results of different distances from the filaments are shown as different colours: blue for $\rm 0 < d_{fil}/\mpchi < 4 $, orange for $\rm 4 < d_{fil}/\mpchi < 8,$ and red for $\rm 8 < d_{fil}/\mpchi < 12$. }
    \label{fig_MS}
\end{figure*}
The \hi-stellar-mass relation is also a commonly studied scaling relation \citep{Saintonge2022}. For \hi observations beyond the local Universe, the 21-cm emission signal becomes increasingly weaker. It is also more efficient to stack \hi spectra for star-forming galaxies to study the \hi-stellar-mass relation at these high redshifts \citep{Chowdhury2022, Sinigaglia2022}. To study the effect of cosmic filaments on the \hi-stellar-mass relation, we focused on the star-forming central galaxies in both observations and models. Due to the resolution limit of the ALFALFA telescope \citep[with a beam size of 3.8 arcmin;][] {Haynes2018}, the \hi mass measurements for satellites are severely affected by the confusion effect, since they typically live in denser environments than those of central galaxies with the same stellar masses. In Figure~\ref{fig_MS}, we show the \hi- and $\rm H_2$-stellar-mass relations for galaxies with different $d_{\rm fil}$ bins in the top and bottom panels, respectively. We compare NUM (solid lines) with TNG100 (dotted lines) in $z = 0$, $z = 0.37$, $z = 1$, and $z = 2$ in the panels from left to right. The bootstrap errors are also shown as shaded areas. 

Our measurements of the \hi-stellar mass relation for star-forming centrals in different $d_{\rm fil}$ bins are shown as the filled circles in panel (a) of Figure~\ref{fig_MS}, with different colours representing different $d_{\rm fil}$ bins. Star-forming galaxies were selected using the criteria of $\log{\rm SFR}>0.65\log M_\ast-7.25$ as in \cite{Guo2021}. The dependence of the \hi-stellar-mass relation on $d_{\rm fil}$ is very weak. We find a weak trend whereby galaxies closer to filaments tend to have slightly smaller $M_{\hj}$ for $M_\ast<10^{10}\msun$. That is, the gas depletion from filaments plays\ a minor role for low-mass galaxies. 

We also include the measurements at $z\sim0.37$ from \cite{Sinigaglia2024} of galaxies within the high- ($\delta>0$, green circle) and low-density ($\delta\leq0$, blue circle) environments, as well as the measurement of the entire sample (red circle). We note that their measurements are for all the star-forming galaxies in different environments, rather than being only for the central galaxies.  

To select the star-forming central galaxies in TNG100, we first used galaxies with $\log (M_\ast/\msun) > 8.5$ to define the star-formation main sequence (SFMS) at each redshift. Following \cite{Ma2022}, we calculated the median values of the SFR in each $M_\ast$ bin for central galaxies and fit the SFR--$M_\ast$ relation with a power law. We define the quenched galaxies using a cut of 1~dex below the SFMS. We iterate the fitting process by removing the quenched galaxies every time until the fits become stable. The final cuts for separating star-forming and quenched central galaxies at different redshifts are as follows:
\begin{eqnarray}
    \log({\rm SFR}_{\rm cut,TNG,z=0}/{\rm yr}^{-1}\msun) &=& 0.86\log M_* - 9.51,\\
    \log({\rm SFR}_{\rm cut,TNG,z=0.5}/{\rm yr}^{-1}\msun) &=& 0.92\log M_* -9.85,\\
    \log({\rm SFR}_{\rm cut,TNG,z=1}/{\rm yr}^{-1}\msun) &=& 0.98\log M_* -9.10,\\
    \log({\rm SFR}_{\rm cut,TNG,z=2}/{\rm yr}^{-1}\msun) &=& 0.99\log M_* -9.85.
\end{eqnarray}
As TNG100 does not have the corresponding \hi measurements at $z=0.37$, we instead used the snapshot of $z=0.5$ in panels (b) and (f). We expect that the weak evolution at low redshifts will not affect the comparisons. In NUM, we adopted a similar definition of $\log{\rm SFR}>\log{\rm SFR_{MS,obs}}-1$ for star-forming central galaxies at each redshift. 

At $z = 0$, TNG100 agrees with observations for low-mass galaxies ($M_\ast<10^{10.5}\msun$), but overpredicts \hi mass for more massive galaxies, consistent with the results of \cite{Ma2022}. This is caused by the fact that the relatively weak kinetic active galactic nucleus (AGN) feedback in TNG100 could only push the cold gas in the inner stellar disc to the outer parts, which is inconsistent with the overall reduction of \hi gas in observation. Similarly to the results shown in Figure~\ref{fig_MH}, the effect of filaments on the \hi-stellar mass relation is relatively weak for low-mass galaxies and becomes stronger for massive ones. Galaxies far from filaments always have lower $M_{\hj}$ at all redshifts. 

The NUM model shows much better agreement with the ALFALFA measurements at $z\sim0$ and reproduces the dependence on $d_{\rm fil}$ for $M_\ast<10^{10}\msun$ exactly. For more massive galaxies, the NUM model also agrees with the observation for $0<d_{\rm fil}/\mpchi<4$. For massive galaxies farther away from the filaments, the NUM model predicts a lower $M_{\hj}$ as in the case of the \hi-halo mass relation. However, no clear dependence on $d_{\rm fil}$ for massive galaxies is seen in observation. This discrepancy is investigated further in Section~\ref{sec:discussion}. 

For fair comparisons with the measurements of \cite{Sinigaglia2024}, in panel (b) we show the model predictions using all star-forming galaxies from NUM and TNG100 with solid and dotted black lines, respectively. The NUM model agrees well with the measured $M_\hj$ in $z\sim0.37$ from \cite{Sinigaglia2024}. TNG100 underestimates the \hi masses for these massive galaxies and predicts a much more significant effect of filament environments. At higher redshifts, the effect of filaments becomes weaker for both the low-mass and high-mass galaxies.

There is no obvious dependence of the H$_2$-stellar-mass relation on the distance to the filaments for both models, as shown in the lower panels of Figure~\ref{fig_MS}. This is expected since the NUM model determines the H$_2$ mass from the stellar mass and the SFR as in Equation~\ref{eq:MH2}. The same behaviour shown in TNG100 further confirms that the H$_2$ mass is not affected by the cosmic web.  

\subsection{Evolution of cold-gas density}\label{subsec:rho} 

\begin{figure*}
    \centering
    \includegraphics[width=0.9\textwidth]{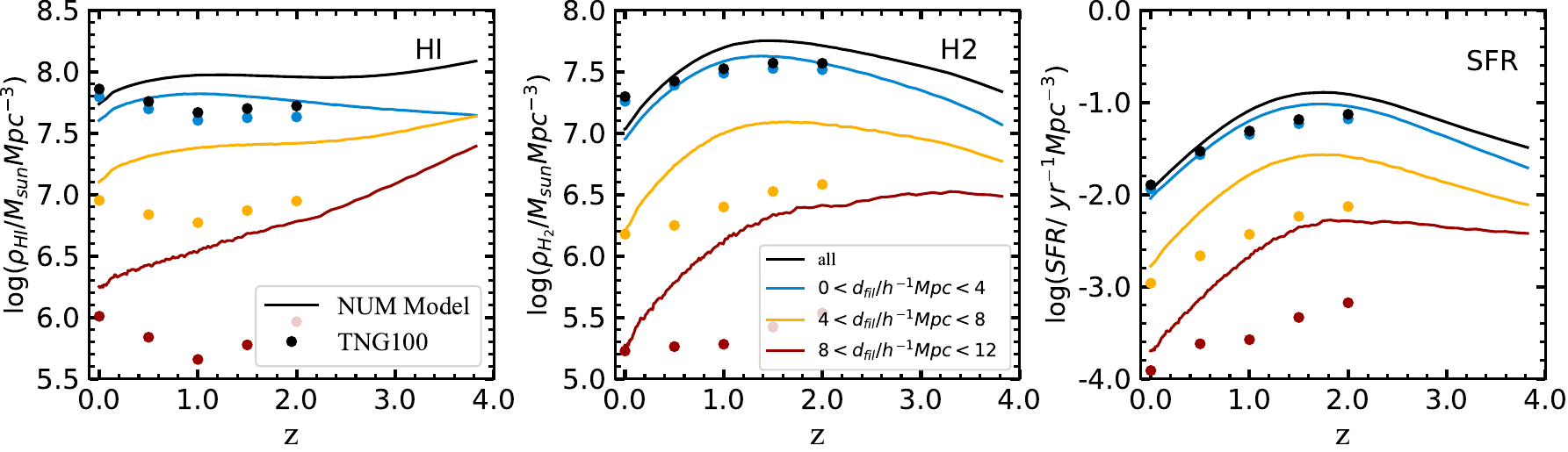}
    \caption{Evolution of \hi (left panel) $\rm H_2$ (middle panel) mass density and SFR (right panel) in NUM model (solid lines) and TNG100 (filled circles). The results of different distance to filaments are shown as different colours: black for all galaxies, blue for $\rm 0 < d_{\rm fil}/\mpchi < 4 $, orange for $\rm 4 < d_{\rm fil}/\mpchi < 8$, and red for $\rm 8 < d_{\rm fil}/\mpchi < 12$.}
    \label{fig_rho_sfr}
\end{figure*}
It is shown in \cite{Wang2024} that the cosmic filaments have undergone a two-phase formation scenario, with a rapid contraction before $z=1$ and a slow growth thereafter. These results are consistent with what we found in the previous sections. The influence of filaments on the \hi gas becomes more significant after $z=1$. It is then intriguing to investigate the evolution of cosmic cold-gas densities in different environments. 

In Figure~\ref{fig_rho_sfr}, we show the evolution of cosmic \hi density $\rho_\hj$ (left panel), cosmic $\rm H_2$ density $\rho_{\rm H_2}$ (middle panel), and SFR density (right panel) for galaxies with different $d_{\rm fil}$ (lines of different colours) in the NUM model. For TNG100, we only have the \hi and H$_2$ measurements in a few snapshots, which are shown as the filled circles. The evolution of $\rho_\hj$ in TNG100 is generally very weak, which is not consistent with the observations at these redshifts \citep{Guo2023}.

In the NUM model, the cosmic \hi and H$_2$ densities are mostly contributed by galaxies close to filaments since $z=4$. The \hi density $\rho_{\hj}$ is increasing with time for galaxies with $0 < d_{\rm fil}/\mpchi < 4$ before $z=1$, while the $\rho_\hj$ for galaxies far from filaments are always decreasing with time. It reflects the fact that the \hi gas in the filaments is accumulating along with the flow of matter to form the filaments before $z=1$. The decrease of $\rho_\hj$ after $z=1$ is likely caused by the consumption of \hi gas and the stellar and AGN feedback mechanisms \citep{Guo2022,Ma2022}. 

The H$_2$ mass density increases before $z\sim1$ and then decreases, closely following the evolution of the SFR density. Galaxies closer to the filaments also tend to have a relatively faster growth of the H$_2$ density before $z=1$, which is due to the accumulation of matter towards the filaments. As discussed in \cite{Wang2024}, mass transport along the filaments after $z=1$ contributes to the formation of clusters that are located at the node structures. We find that galaxies within $2\mpchi$ of the nodes contribute $\sim20\%$ of the total $\rho_\hj$ and $\sim31\%$ of the total $\rho_{\rm H_2}$ at $z<1$. The corresponding fractions at $z=4$ are around $3\%$ and $13\%$ for \hi and H$_2$, respectively. It is consistent with the scenario suggested in \cite{Wang2024} that cluster formation is supported by the cold-gas flow along the filaments to the nodes at $z<1$. 

\section{Discussion}\label{sec:discussion}
In previous sections, we find that the \hi gas is mainly determined by halo mass for low-mass haloes and becomes more sensitive to filaments for massive haloes. This change occurs at $\log(M_{\rm vir}/\msun)\sim12.5$, where the trend of \hi-halo mass relation with different $d_{\rm fil}$ also reverses. The depletion of \hi gas in low-mass haloes when approaching the filaments seems to dominate over the accretion of \hi gas in denser environments. This behaviour is seen in both the observations and the NUM model, but not in the TNG100 simulation. We discuss what causes the reverse of the \hi -halo mass and \hi-stellar-mass relations in Section \ref{subsec:dis_tranision} and compare our results with the literature in Section \ref{subsec:dis_compare}.

\begin{figure*}
    \centering
    \includegraphics[width=\textwidth]{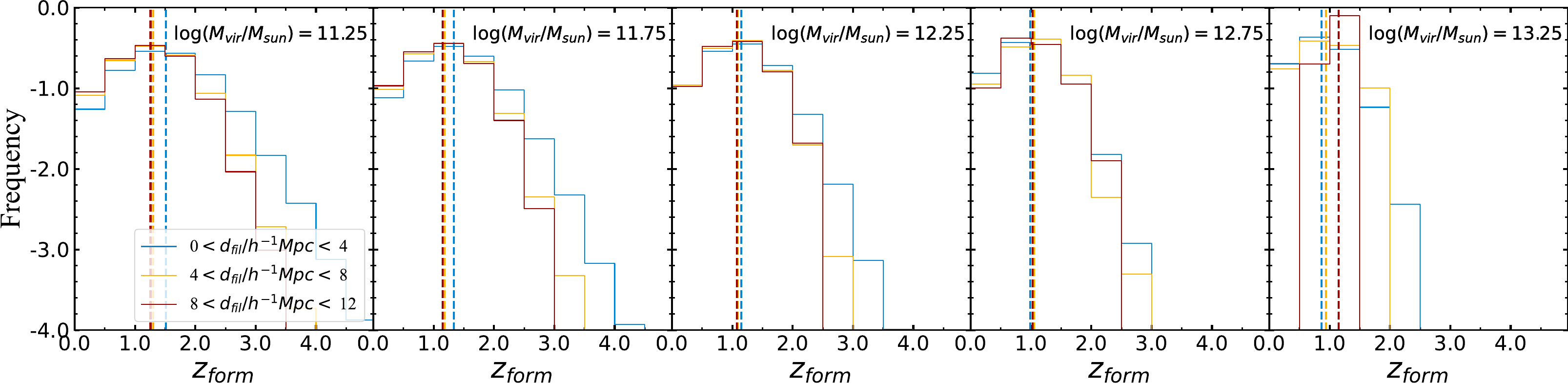}
    \caption{Distribution of halo-formation time ($z_{\rm form}$) in NUM model. From left to right we show the distribution of haloes with $\log (M_{\rm vir}/\msun) = 11.25, 11.75, 12.25, 12.75, 13.25$, respectively. We show the mean $z_{\rm form}$ for each population as a vertical dashed line. Blue, orange, and red ones represent galaxies with $\rm 0 < d_{fil}/\mpchi < 4$, $\rm 4 < d_{fil}/\mpchi < 8$, $\rm 8 < d_{fil}/\mpchi < 12$, respectively.}
    \label{fig_HIMH_reverse}
\end{figure*}

\begin{figure*}
    \centering
    \includegraphics[width=0.8\textwidth]{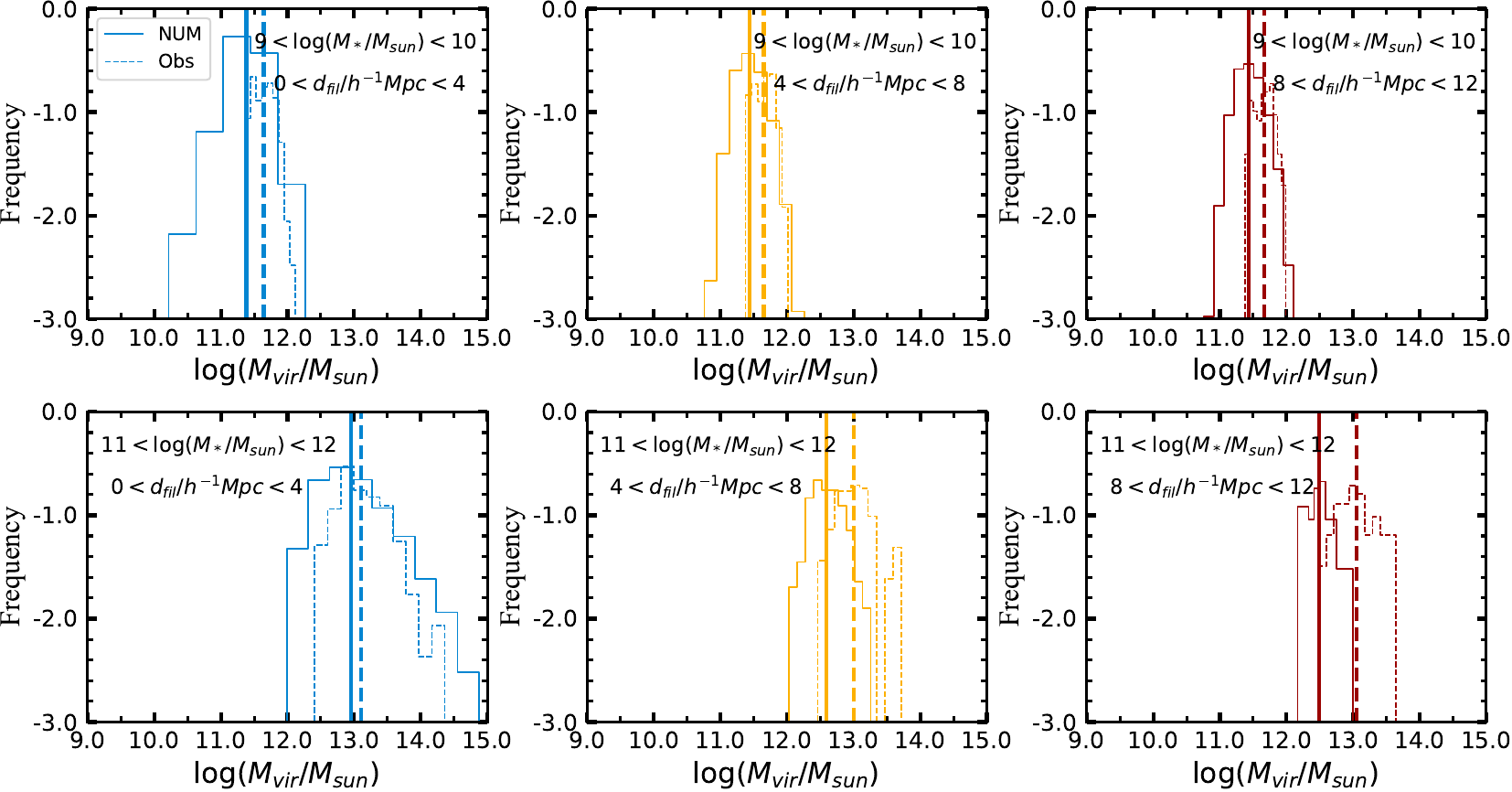}
    \caption{Comparison of halo-mass distribution for star-forming central galaxies in NUM model (solid lines) and observations (dashed lines) at $z=0$. We show the mean halo masses for each population as vertical lines, solid lines are used for the NUM model, and dashed lines are used for observations. The top row shows galaxies with $9<\log(M_\ast/\msun)<10$, and the bottom row shows galaxies with $11<\log(M_\ast/\msun)<12$. From left to right, we show the distribution of galaxies with $\rm 0 < d_{fil}/\mpchi < 4 $ (left panels), $\rm 4 < d_{fil}/\mpchi < 8$ (middle panels), and $\rm 8 < d_{fil}/\mpchi < 12$ (right panels), respectively.}
    \label{fig_HIMS_reverse}
\end{figure*}

\subsection{Reversed trends in the \hi-halo-mass and \hi-stellar-mass relations}\label{subsec:dis_tranision}
In the NUM model, $M_{\rm \hj}$ is completely determined by $M_{\rm vir}$, $z_{\rm form}$, and SFR at each redshift, as determined in Equation~\ref{eq:MHI}. This means that the \hi-halo-mass relation is likely to be only affected by the different $z_{\rm form}$ and SFR distributions in different $d_{\rm fil}$ environments. We checked that the SFR-halo-mass distributions in different $d_{\rm fil}$ bins are quite similar to each other. Thus, the different trends of the \hi-halo-mass relation with $d_{\rm fil}$ could only be caused by variations in the $z_{\rm form}$ parameter. 

In Figure~\ref{fig_HIMH_reverse}, we show the distributions of $z_{\rm form}$ in different halo-mass bins for the NUM model. The measurements in different $d_{\rm fil}$ bins are distinguished by the colours. From left to right, we show the distributions of $z_{\rm form}$ for haloes with the median masses of $\log(M_{\rm vir}/\msun) = 11.25, 11.75, 12.25, 12.75, 13.25$, respectively. The mean values of $z_{\rm form}$ in different $d_{\rm fil}$ bins are displayed as vertical dashed lines of different colours. We can see that, on average, low-mass haloes ($\log (M_{\rm vir}/\msun)<12.5$) close to filaments ($0 < d_{\rm fil}/\mpchi < 4$) have higher $z_{\rm form}$ (i.e. formed early) than those far from filaments. Those haloes with $4 < d_{\rm fil}/\mpchi< 8$ and $8 < d_{\rm fil}/\mpchi< 12$ always have similar $z_{\rm form}$ values. 

However, for more massive haloes ($\log (M_{\rm vir}/\msun)>12.5$), those closer to the filaments would have lower $z_{\rm form}$. The difference in the $z_{\rm form}$ values of different $d_{\rm fil}$ bins increases with the halo mass. As Equation~\ref{eq:MHI} shows, the haloes that form late will maintain larger $M_{\rm \hj}$. Thus, the variation of $z_{\rm form}$ in different $d_{\rm fil}$ environments explains the reverse of the trend with $d_{\rm fil}$ in the \hi-halo and \hi-stellar-mass relations. At higher redshifts, the dependence of $M_\hj$ on $z_{\rm form}$ becomes weaker ($M_{\hj} \propto (1+z_{\rm form})^{-\gamma}$), leading to reduced effects of filaments. 

Physically, low-mass haloes grow most of their mass through smooth accretion from the environment \citep{Genel2010}. Those located close to filaments will thus form earlier with efficient mass accretion. They would also suffer from sustained \hi gas depletion in such a dense environment. On the other hand, the growth of massive haloes with $\log(M_{\rm vir}/\msun)>12.5$ is dominated by mergers \citep[see e.g. Fig.~5 of][]{Genel2010}. Massive haloes closer to filaments would experience many more merger events and thus have later formation times. The \hi gas brought in by the mergers would compensate for the strong depletion in the filament environment. It also explains the effect that removing the haloes within $d_{\rm node}<2\mpchi$ would result in a flat \hi-halo-mass relation at the massive end, since the haloes in the nodes would have even later formation times. Moreover, $M_{\hj,{\rm tot}}$ in massive haloes would have more contributions from the satellite galaxies. The haloes in denser environments would have more satellite galaxies, which also increases the total \hi masses \citep{Guo2020}.

In the \hi-stellar-mass relation of Figure~\ref{fig_MS}, we find that the NUM model is consistent with the observation for low-mass galaxies, but it under-predicts the \hi masses for massive galaxies with $d_{\rm fil}>4\mpchi$. Given the good agreement in the \hi -halo-mass relation, such a discrepancy is probably caused by the different halo-mass distributions. In Figure~\ref{fig_HIMS_reverse}, we show the halo-mass distributions in the NUM model (solid lines) and the observation sample (dotted lines) for the star-forming central galaxies. The mean halo mass for each sample is shown as the vertical line in each panel. In low-mass galaxies of $9<\log(M_\ast/\msun)<10$, the mean values of $M_{\rm vir}$ between the observation and the NUM model are consistent with each other, with the observed one slightly higher by around 0.2~dex. This is caused by the cut-off halo mass around $10^{11.4}\msun$ in the SDSS group catalogue \citep{Lim2017}, which is also evidenced by the narrower distributions of the halo mass in observation. For massive galaxies of $11<\log(M_\ast/\msun)<12$, galaxies close to filaments have consistent halo-mass distributions in observations and in NUM. However, galaxies far from filaments in observation have much higher mean halo masses by around 0.5~dex than the NUM model predictions. This could be caused by the potential central-satellite mis-identification in the group catalogue \citep{Campbell2015}. The satellite galaxies with the same $M_\ast$ as the centrals are typically located in much more massive haloes. Since massive haloes far from filaments have a low number of satellites, the confusion between central and satellite galaxies would become severe for these haloes. 

\subsection{Different models between TNG and NUM}\label{subsec:difference}

The different model predictions of the \hi and H$_2$ gas properties between NUM and TNG provide important clues to understand the evolution of cold gas. We note that in the NUM model, the \hi mass is determined only by $M_{\rm vir}$, $z_{\rm form,}$ and SFR, without including any parameters directly related to the cosmic web. The influence of the cosmic web is coupled with that of the local halo environment. By investigating the \hi-halo-mass relation, we can efficiently separate the effect of halo mass from that of the cosmic filaments. 
 
As seen in the above section, the secondary dependence of the \hi mass on $d_{\rm fil}$, in addition to the dependence on the halo mass and stellar mass, can be largely explained by the influence of the halo-formation time. It was also found in \cite{Stiskalek2021} that populating \hi-rich galaxies in late-forming haloes is necessary to explain the observed \hi clustering measurements. This is consistent with the finding of \cite{Guo2020} that haloes with the same masses but more satellite galaxies tend to have higher \hi masses. Haloes with later formation times tend to undergo more mergers and thus have more satellite members and more \hi gas. Such differences in the halo merger histories are caused by the influence of the cosmic-web environment. We verified that in both NUM and TNG100, the halo-mass distributions as a function of $d_{\rm fil}$ are almost the same. If NUM had a different relation of $M_\hj(M_{\rm vir}, d_{\rm fil})$ from the observation, the resulting \hi-halo mass relation would not be consistent with the observation. 
 
However, the galaxy formation-model parameters in TNG100 are not calibrated against the \hi and H$_2$ observations. The postprocessing framework of TNG decomposes neutral hydrogen into \hi and $\rm H_2$ components -- by modelling the UV radiation background and the ISM density -- which are also not directly related to the properties of the cosmic web. In \cite{Li2022}, it is found that the \hi mass in TNG100 is strongly correlated with the halo spin parameter and shows a weak dependence on the halo formation time. In other words, the \hi mass in TNG100 is more affected by the dynamical property than by the accretion history of the halo. 
 
 As detailed in \cite{Liu2025}, the cold gas in massive galaxies of TNG100 typically has a high angular momentum due to the mechanism of the kinetic AGN feedback. Even quenched galaxies can retain a large amount of \hi gas due to the high angular momentum. Together with the mild energy release of the kinetic AGN feedback, the \hi gas is accumulated in the circumgalatic medium and further leads to the overestimation of \hi mass compared to the observation for massive galaxies. This also explains the strong correlation between the \hi mass and the halo-spin parameter in the TNG, which makes the dependence of the \hi mass on the cosmic web less realistic.

\subsection{Comparison with literature}\label{subsec:dis_compare}
The discrepancies identified in previous studies can be addressed by our findings. \cite{Odekon2018} examined late-type galaxies from the ALFALFA survey and discovered that the \hi mass of galaxies with $8.5<\log(M_\ast/\msun)<10.5$ decreases as they approach filaments. In contrast, \cite{Kleiner2017} used \hi data from the HIPASS survey to show that galaxies with $\log(M_\ast/\msun)>11$ exhibit a consistently higher \hi fraction when nearer to filaments. These observations align well with our findings shown in Figure \ref{fig_MS}. The \hi mass is influenced by the interplay between gas depletion and accretion. Our results suggest that for low-mass galaxies in filaments, gas depletion prevails over smooth accretion, while the opposite holds for massive galaxies. In addition, \cite{Hoosain2024} noted that the gas-poor ($M_\hj/M_\ast<0.1$) fraction consistently rises as galaxies with $8.5<\log(M_\ast/\msun)<11.5$ move closer to the filaments in the RESOLVE and ECO samples. It is important to mention that their data include both central and satellite galaxies, with gas depletion significantly more pronounced in satellite galaxies within dense environments \citep{Brown2017}. Considering all galaxies, our findings also agree with their observations.

Based on the hydrodynamical SIMBA simulation, \cite{Bulichi2024} demonstrated that the \hi fraction of galaxies located in haloes with $\log(M_{\rm vir}/\msun)<13$ decreases within $1\mpchi$ of the filaments, while the H$_2$ fraction remains largely unaffected by their distances to the filaments. This aligns well with our predictions for the NUM model as observed in the \hi-halo- and \hi-stellar-mass correlations. Prior research has led to debates over whether filaments can decrease or increase the gas supply of galaxies. Our findings indicate that this relationship is not linear. Both the local halo environment and the large-scale filaments influence the balance between gas depletion and accretion, with the halo environment having a predominant impact on the cold gas mass. Ongoing and future \hi surveys \citep[e.g.][]{Koribalski2020,Zhang2024,Ma2024} can provide more constraints on the influence of the cosmic web.

\section{Conclusions}\label{sec:conclusion} 

In this study, we investigated the influence of cosmic filaments and the local halo environment on the cold-gas scaling relations using the empirical model NUM and the hydrodynamical simulation TNG100. We compared the model predictions with the measured \hi -halo- and \hi -stellar-mass relations at $0<z<0.05$ using the ALFALFA survey. Our main conclusions are as follows.

\begin{enumerate}
\item Distances to the filaments have little impact on the shapes of the \hi and $\rm H_2$ mass functions and only reduce the total amount of cold gas in both NUM and TNG100. The NUM model shows good agreement with the observed HIMF and H$_2$MF at $z=0$, while TNG100 overestimates the H$_2$MF and the massive end of the HIMF.   

\item The role of filaments in affecting the \hi-halo and \hi-stellar mass relations is generally less significant compared to the halo environment. There is a slight trend where low-mass haloes with $\log(M_{\rm vir}/\msun)<12.5$ that are closer to filaments tend to have reduced \hi masses. However, this pattern is inverted for massive haloes with $\log(M_{\rm vir}/\msun)>12.5$. This behaviour observed in the ALFALFA data is accurately reflected in the NUM model, but it does not appear in the TNG100 simulation. In general, TNG100 forecasts higher \hi masses for haloes and galaxies that are closer to filaments. The reverse pattern is caused by the formation time dependence of the \hi gas in the NUM model and successfully explains the observed behaviour.

\item In both the NUM model and the TNG100 simulation, the H$_2$ mass of a galaxy is mainly influenced by its stellar mass and SFR. The large-scale filament environment does not significantly impact it. The correlation between the H$_2$-halo-mass relation and the distance to filaments for massive haloes is due to the variation of stellar-mass distributions across different filament environments.

\item TNG100 and NUM have very different predictions for the \hi-halo and \hi-stellar-mass relations beyond the local Universe. The NUM model agrees well with the stacked \hi observations at $z\sim0.37$ from \cite{Sinigaglia2024}. However, TNG100 predicts much lower \hi gas masses at these high redshifts. 

\item The cosmic \hi and H$_2$ densities are mainly contributed by galaxies that lie close to the filaments. The formation of filaments is accompanied by the aggregation of cold gas towards the filaments. 
\end{enumerate}

\begin{acknowledgements}
We thank the anonymous reviewer for the helpful comments that improve the presentation of this paper. 
This work is supported by the National SKA Program of China (grant No. 2020SKA0110100), the CAS Project for Young Scientists in Basic Research (No. YSBR-092) and GHfund C(202407031909). We thank Wei Wang for providing the distance to filament measurements in the NUM catalogue. We acknowledge the use of the High Performance Computing Resource in the Core Facility for Advanced Research Computing at the Shanghai Astronomical Observatory.
\end{acknowledgements}

\bibliographystyle{aa}
\bibliography{ref}

\end{document}